\documentclass[11pt,a4paper]{article}
\usepackage{jheppub}
\usepackage{graphicx}
\usepackage{epstopdf}
\usepackage{subfigure}
\usepackage{textcomp}
\def\bkR{{\rm I\kern-.17em R}}
\def\bkC{{\rm \kern.24em \vrule width.05em height1.4ex depth-.05ex \kern-.26em C}}
\def\to{\rightarrow}

\def\JHEP{{\it JHEP} }

\def\be{\beta}

\def\frac#1#2{{\textstyle{{#1}\over {#2}}}}

\def\lsim{\mathrel{\rlap{\lower4pt\hbox{\hskip1pt$\sim$}}
    \raise1pt\hbox{$<$}}}
\def\gsim{\mathrel{\rlap{\lower4pt\hbox{\hskip1pt$\sim$}}
    \raise1pt\hbox{$>$}}}
\def\sqr#1#2{{\vcenter{\vbox{\hrule height.#2pt
         \hbox{\vrule width.#2pt height#1pt \kern#1pt
         \vrule width.#2pt}
         \hrule height.#2pt}}}}

\def\laq{\raise 0.4 ex \hbox{$<$}\kern -0.8 em\lower 0.62 ex\hbox{$\sim$}}
\def\gaq{\raise 0.4 ex \hbox{$>$}\kern -0.7 em\lower 0.62 ex\hbox{$\sim$}}

\def\be{\begin{equation}}
\def\ee{\end{equation}}
\def\ba{\begin{eqnarray}}
\def\ea{\end{eqnarray}}

\def\dalemb#1#2{{\vbox{\hrule height.#2pt
        \hbox{\vrule width.#2pt height#1pt \kern#1pt \vrule width.#2pt}
        \hrule height.#2pt}}}

\def\dalemb#1#2{{\vbox{\hrule height.#2pt
        \hbox{\vrule width.#2pt height#1pt \kern#1pt \vrule width.#2pt}
        \hrule height.#2pt}}}

\def\gtorder{\mathrel{\raise.3ex\hbox{$>$}\mkern-14mu
             \lower0.6ex\hbox{$\sim$}}}
\def\ltorder{\mathrel{\raise.3ex\hbox{$<$}\mkern-14mu
             \lower0.6ex\hbox{$\sim$}}}

\title{Non-Canonical Phase-Space Noncommutativity and the Kantowski-Sachs singularity for Black Holes}

\author[a]{Catarina Bastos}

\author[a,b]{Orfeu Bertolami}

\author[c,d]{Nuno Costa Dias} 

\author[c,d]{and Jo\~ao Nuno Prata}

\affiliation[a]{Instituto de Plasmas e Fus\~ao Nuclear, Instituto Superior T\'ecnico, \\
Avenida Rovisco Pais 1, 1049-001 Lisboa, Portugal}

\affiliation[b]{Departamento de F\'\i sica e Astronomia, \\
Faculdade de Ci\^encias da Universidade do Porto \\
Rua do Campo Alegre, 687,4169-007 Porto, Portugal}

\affiliation[c]{Departamento de Matem\'{a}tica, Universidade Lus\'ofona de Humanidades e Tecnologias,\\
 Avenida Campo Grande, 376, 1749-024 Lisboa, Portugal}

\affiliation[d]{Grupo de F\'{\i}sica Matem\'atica, UL,\\
Avenida Prof. Gama Pinto 2, 1649-003, Lisboa, Portugal. }

\emailAdd{catarina.bastos@ist.utl.pt}

\emailAdd{orfeu.bertolami@fc.up.pt}

\emailAdd{ncdias@meo.pt}

\emailAdd{joao.prata@mail.telepac.pt}

\abstract{We consider a cosmological model based upon a non-canonical noncommutative extension of the Heisenberg-Weyl algebra to address the thermodynamical stability and the singularity problem of both the Schwarzschild and the Kantowski-Sachs black holes. The interior of the black hole is modelled by a noncommutative extension of the Wheeler-DeWitt equation. We compute the temperature and entropy of a Kantowski-Sachs black hole and compare our results with the Hawking values. Again, the noncommutativity in the momenta sector allows us to have a minimum in the potential, which is relevant in order to apply the Feynman-Hibbs procedure. For Kantowski-Sachs black holes, the same model is shown to generate a non-unitary dynamics, predicting vanishing total probability in the neighborhood of the singularity. This result effectively regularizes the Kantowski-Sachs singularity and generalizes a similar result, previously obtained for the case of Schwarzschild black hole.}

\begin{document}

\maketitle

\section{Introduction}

Noncommutativity has been recurrently considered as a fundamental
feature of space-time at the Planck scale \cite{CS}. Indeed, it is
advocated that noncommutative geometry (NCG) captures the fuzzy
nature of space-time at this scale, and that it should emerge as
one of the main effects of quantum gravity \cite{Connes,Madore}.
In the context of string and M-theory, a configuration space
noncommutativity arises as the low-energy effective theory of a
D-brane in the background of a Neveu-Schwarz B field living in a
space with spatial noncommutativity \cite{CS}. More recently,
various aspects of noncommutative quantum field theory
\cite{DS,OBCZ} and quantum mechanics \cite{Bastos1,OB,ZA,Bastos9} have been
investigated.

Noncommutativity has also been considered in the context of Black
Holes (BH) to address the issues of thermodynamical stability and
the removal/regularization of singularities. In this framework, we
have argued in a series of papers that an additional
momentum-momentum noncommutativity leads to interesting and
relevant properties \cite{Bastos2,Bastos5,Bastos3}\footnote{The impact of configuration space noncommutativity on BH has been throughly reviewed in Ref. \cite{Nicolini}}. In these works
the BH is modelled with the Kantowski-Sachs (KS) metric \cite{Kantowski} and,
through the ADM procedure, the corresponding Wheeler-de Witt
(WDW) equation is obtained and solved. In Refs. \cite{Bastos2,Bastos5} we have considered
a canonical phase-space noncommutativity for the two scale factors
and the conjugate momenta of the KS geometry. There are two
noteworthy consequences of this choice of the algebra. To start with,
the solution of the WDW equation factorizes into an oscillatory
part of a ``time" variable and a ``spatial" part which obeys an
ordinary Schr\"odinger-like differential equation. The associated
potential of this equation exhibits a local minimum. This allows
for an expansion around the minimum and the saddle point
evaluation of the partition function in accordance with the
Feynman-Hibbs procedure \cite{Feynman}. From the rules of
statistical mechanics, we can then compute the temperature and entropy of
the BH. The results are tantamount to the Hawking-Bekenstein
quantities plus stringy and noncommutative corrections. However,
the essential point is that the local minimum of the potential
only appears if momenta noncommutativity is assumed. The second
consequence of the noncommutative algebra is that the ``spatial" part of the wave
function displays a pronounced damping behavior, and shown, that in fact, 
vanishes at infinity \cite{Bastos2}. However, the obtained rate of
decay is not sufficient to ensure square-integrability of
the wave-function. But this is only marginally so. This prompted
the search for alternative noncommutative algebras which might
lead to a full-fledged square-integrable wave-function and the
corresponding regularization of the BH singularity. In Ref.
\cite{Bastos3} we have proposed a non-canonical phase-space
noncommutative algebra, which delivers a normalizable ``spatial"
wave function. This algebra is a parsimonious extension of the
Heisenberg-Weyl (HW) algebra in the sense that: (i) it includes
both position-position and momentum-momentum noncommutativity;
(ii) it only requires one extra noncommutative parameter; (iii) it
is globally isomorphic with the Heisenberg-Weyl algebra. In Ref.
\cite{Bastos3} we have shown that this algebra effectively
regularizes the singularity of the Schwarzshild BH at the quantum
level.

In the present work we further investigate the implications of
this algebra. We compute the thermodynamical quantities of the BH,
which are shown to have extra non-canonical noncommutative
corrections. Moreover, we prove that this algebra also regularizes
the singularity of the KS BH.

This paper is organized as follows. In section 2, we review the
main aspects of the non-canonical noncommutative algebra of Ref.
\cite{Bastos3} and derive the Schr\"odinger-type equation
satisfied by the ``spatial" part of the wave function. In section
3, we evaluate the partition function of the BH by the
Feynman-Hibbs method and compute the temperature and entropy of
the BH. In section 4, we study the probability of the BH near the
KS singularity. Finally, in section 5, we discuss some of the
mathematical subtleties of the regularization result and present
our conclusions. In this work we use units $\hbar=c=G=1$.

\section{Phase-space non-canonical noncommutative quantum cosmology}

The Schwarzschild BH is described by the metric, \be\label{eq0.1}
ds^2=-\left(1-{2M\over r}\right)dt^2+\left(1-{2M\over
r}\right)^{-1}dr^2+r^2 d\Omega^2~, \ee where $r$ is the radial
coordinate and $d\Omega^2=d\theta^2+\sin^2\theta d\varphi^2$. For
$r<2M$ the time and radial coordinates are interchanged
($r\leftrightarrow t$) so that space-time is described by the
metric: \ba\label{eq0.2} ds^2=-\left({2M\over
t}-1\right)^{-1}dt^2+\left({2M\over t}-1\right)dr^2+t^2
d\Omega^2~. \ea This is an anisotropic metric, thus for $r<2M$,
the interior of a Schwarzschild BH can be described as an
anisotropic cosmological space-time. Indeed, the metric
(\ref{eq0.2}) can be mapped into the KS metric \cite{Dominguez},
which, in the Misner parameterization, can be written as
\be\label{eq0.3}
ds^2=-N^2dt^2+e^{2\sqrt{3}\beta}dr^2+e^{-2\sqrt{3}\beta}e^{-2\sqrt{3}\Omega}d\Omega^2~,
\ee where $\Omega$ and $\beta$ are scale factors, and $N$ is the
lapse function. The following identification for $r<2M$
\cite{Bastos5,Bastos3}: \be\label{eq0.4} N^2=\left({2M\over
t}-1\right)^{-1}
\hspace{0.2cm},\hspace{0.2cm}e^{2\sqrt{3}\beta}=\left({2M\over
t}-1\right)\hspace{0.2cm},\hspace{0.2cm}e^{-2\sqrt{3}\beta}e^{-2\sqrt{3}\Omega}=t^2~.
\ee is a surjective mapping that transforms the metric Eq.
(\ref{eq0.3}) into the metric Eq. (\ref{eq0.2}). Moreover, we also
have from (\ref{eq0.4}) that 
\be\label{eq0.5}
e^{-2\sqrt{3}\Omega}=t(2M-t) 
\ee
and so the Schwarzschild
singularity ($t \to 0$) corresponds to \be\label{eq0.6}
t\longrightarrow 0^+ \Longrightarrow \Omega,\beta \longrightarrow
+\infty. \ee On the other hand, the KS singularity is attained at
(cf. Eq.(\ref{eq0.3})) $\beta \to +\infty$.

In this paper we shall study the thermodynamics of the interior of
the black hole by considering the following non-canonical
extension of the HW algebra \cite{Bastos3}: \ba\label{eq1.5}
\left[\hat{\Omega}, \hat{\beta} \right]\! &=&\! i \theta \left( 1 + \epsilon \theta \hat{\Omega} + {\epsilon \theta^2\over{1 + \sqrt{1- \xi}}} \hat{P}_{\beta} \right)\\
\left[\hat{P}_{\Omega}, \hat{P}_{\beta} \right]\! &=&\! i \left( \eta  + \epsilon (1 + \sqrt{1 - \xi})^2 \hat{\Omega} + \epsilon \theta (1 + \sqrt{1- \xi}) \hat{P}_{\beta} \right)\nonumber\\
\left[\hat{\Omega}, \hat{P}_{\Omega} \right]\! &=&\! \left[\hat{\beta}, \hat{P}_{\beta} \right]\! =\!i  \left( 1 + \epsilon \theta (1 + \sqrt{1-
\xi}) \hat{\Omega} + \epsilon \theta^2 \hat{P}_{\beta} \right)\nonumber,
\ea
where $\theta$, $\eta$ and $\epsilon$ are positive constants and $\xi = \theta \eta <1$. Notice that this condition is experimentally satisfied in at least two distinct situations \cite{OB,Bastos10}. 
The remaining commutation relations vanish. For $\epsilon\neq0$ it implies that the noncommutative commutation and
uncertainty relations are themselves position and momentum dependent. Notice that $\epsilon=0$ corresponds to the canonical phase-space noncommutativity \cite{Bastos1,Bastos2,Bastos5}.
We point out that the well known Darboux's Theorem ensures that we can always find a system of coordinates, where locally the algebra can be written as the HW algebra. For this algebra, this statement also holds globally, so that the algebra is isomorphic to the HW algebra. Actually, algebra Eqs. (\ref{eq1.5}) is the simplest non-canonical extension that is globally isomorphic to the HW algebra.  It is an open issue whether this algebraic structure can be physically motivated or derived from a fundamental theory. On the other hand, this non-canonical extension has a direct impact on the singularity problem as shown in Ref. \cite{Bastos3}.

The isomorphism between the algebra Eqs. (\ref{eq1.5}) and the HW
algebra is referred to as a Darboux (D) transformation. D
transformations are not unique. Indeed, the composition of a D
transformation with a unitary transformation yields another D
transformation. However, the physical predictions are invariant
under different choices of the D map. Suppose that
$\left(\hat{\Omega}_c, \hat{P}_{\Omega_c}, \hat{\beta_c},
\hat{P}_{\beta_c} \right)$ obey the HW algebra: \be\label{eq1.11}
\left[\hat{\Omega}_c, \hat{P}_{\Omega_c} \right] =
\left[\hat{\beta}_c, \hat{P}_{\beta_c} \right] =i~. \ee The
remaining commutation relations vanish. Then a suitable
transformation is: \ba\label{eq1.7}
\hat{\Omega} \!\!&=&\!\! \lambda \hat{\Omega}_c -  {\theta\over2\lambda} \hat{P}_{\beta_c} + E \hat{\Omega}_c^2\hspace{0.2cm},\hspace{0.2cm}\hat{\beta} = \lambda \hat{\beta}_c + {\theta\over2 \lambda} \hat{P}_{\Omega_c} \nonumber\\
\hat{P}_{\Omega}\!\! &=&\!\! \mu \hat{P}_{\Omega_c} +
{\eta\over2\mu} \hat{\beta}_c
\hspace{0.2cm},\hspace{0.2cm}\hat{P}_{\beta} = \mu
\hat{P}_{\beta_c} -  {\eta\over2 \mu} \hat{\Omega}_c + F
\hat{\Omega}_c^2. \ea Here, $\mu$, $ \lambda$ are real parameters
such that $(\lambda \mu)^2 - \lambda \mu + \frac{\xi}{4}
=0\Leftrightarrow 2 \lambda \mu = 1 \pm \sqrt{1- \xi}$, and we
choose the positive solution (given the invariance of the physics
under different choices of the D map \cite{Bastos1}), and set
\be\label{eq1.9} E = - {\theta\over{1 + \sqrt{1- \xi}}}
F\hspace{0.2cm},\hspace{0.2cm}F = - {\lambda\over\mu} \epsilon
\sqrt{1- \xi} \left(1 + \sqrt{1- \xi} \right)~. \ee The inverse D
map is easily computed: \ba\label{eq1.10}
\hat{\Omega}_c &=& {1\over\sqrt{1- \xi}} \left( \mu \hat{\Omega} + {\theta\over2 \lambda} \hat{P}_{\beta}  \right)\nonumber\\
\hat{P}_{\Omega_c} &=& {1\over\sqrt{1- \xi}} \left(\lambda \hat{P}_{\Omega} - {\eta\over2 \mu} \hat{\beta} \right)\nonumber\\
\hat{\beta}_c &=& {1\over\sqrt{1- \xi}} \left(\mu \hat{\beta} - {\theta\over2\lambda} \hat{P}_{\Omega} \right) \\
\hat{P}_{\beta_c} &= & {1\over\sqrt{1- \xi}}\left[\lambda
\hat{P}_{\beta} +  {\eta\over2 \mu} \hat{\Omega} -{F
\mu\over\sqrt{1- \xi}} \left( \hat{\Omega} +
{\theta\over{1+\sqrt{1- \xi}}} \hat{P}_{\beta}
\right)^2\right]\nonumber. \ea It can be shown using Eqs.
(\ref{eq1.7}) and (\ref{eq1.10}) that the algebra Eq.
(\ref{eq1.5}) implies the HW algebra, Eq. (\ref{eq1.11}), and
vice-versa. Of course, as already pointed out, we could consider
other quadratic transformations relating the two algebras, which
would, nevertheless, lead to the same physical predictions
\cite{Bastos1}.

We consider now the Hamiltonian associated to the WDW equation for the KS metric and a particular factor ordering \cite{Bastos2}
\be\label{eq1.12}
\hat H = - \hat{P}_{\Omega}^2 + \hat{P}_{\beta}^2 - 48 e^{-2 \sqrt{3} \hat{\Omega}}.
\ee
Upon substitution of Eqs. (\ref{eq1.7}), we obtain:
\ba\label{eq1.13}
\hat H&=& - \mu^2 \hat{P}_{\Omega_c}^2 - {\eta^2\over4 \mu^2} \hat{\beta}_c^2 - \eta \hat{\beta}_c \hat{P}_{\Omega_c} + \mu^2 \hat{P}_{\beta_c}^2 + {\eta^2\over4 \mu^2} \hat{\Omega}_c^2 +F^2 \hat{\Omega}_c^4 - \eta \hat{\Omega}_c \hat{P}_{\beta_c}\nonumber\\
&+& 2 \mu F \hat{\Omega}_c^2 \hat{P}_{\beta_c} - {\eta F\over\mu}
\hat{\Omega}_c^3-48 \exp \left( -2 \sqrt{3} \lambda \hat{\Omega}_c
+ {\sqrt{3} \theta\over\lambda}\hat{P}_{\beta_c}- 2 \sqrt{3} E
\hat{\Omega}_c^2 \right)~. \ea In the previous expression, the HW
operators have the usual representation: $\hat{\Omega}_c =
\Omega_c$, $\hat{\beta}_c = \beta_c$, $\hat{P}_{\Omega_c}= -i
\frac{\partial}{\partial \Omega_c}$ and $\hat{P}_{\beta_c}= -i
\frac{\partial}{\partial \beta_c}$. Let us now define the operator
$\hat A$, which corresponds to a constant of motion of the
classical problem as discussed in Ref. \cite{Bastos2} (which
considers the same space-time setup, but assumes a canonical
phase-space noncommutative algebra): \be\label{eq1.14} \hat A =
\mu  \hat{P}_{\beta_c} + {\eta\over2 \mu} \hat{\Omega}_c. \ee A
simple calculation reveals that for the noncommutative algebra Eq.
(\ref{eq1.5}), $\hat A$ also commutes with the Hamiltonian $\hat
H$, given by Eq. (\ref{eq1.12}), i.e. $[\hat H, \hat A]=0$. This
quantity corresponds to the momentum $P_{\beta}$ shifted by
$(\eta/(2\mu^2))\Omega$, which can be seen as analogous to the
canonical conjugate momentum in the presence of a gauge field,
where $\eta/2\mu^2$ corresponds to the electric charge and
$\Omega$ to the gauge field component.

Thus, one seeks for solutions $\psi \left( \Omega_c, \beta_c
\right)$ of the WDW equation, \be\label{eq1.16} \hat H \psi = 0~,
\ee which are simultaneous eigenstates of $\hat A$. The most
general solution of the eigenvalue equation $\hat A \psi = a \psi$
with $a$ real is \cite{Bastos5,Bastos3} \be\label{eq1.18} \psi_a
\left( \Omega_c, \beta_c \right)= {\cal R} \left( \Omega_c \right)
\exp \left[ {i \beta_c\over\mu} \left(a - {\eta\over2 \mu}
\Omega_c \right) \right], \ee where ${\cal R} \left(\Omega_c
\right)$ is an arbitrary $C^2$ function of $\Omega_c$, which is
assumed to be real.

>From Eqs. (\ref{eq1.13}), (\ref{eq1.16}), and (\ref{eq1.18}), one gets after some algebraic manipulation
\ba\label{eq1.19}
&&\mu^2 {\cal R}''- 48 \exp \left( - {2 \sqrt{3}\over\mu} \Omega_c - 2 \sqrt{3} E \Omega_c^2 + {\sqrt{3} \theta a\over\mu \lambda} \right) {\cal R}- {2 \eta\over\mu} \left(\Omega_c+F\Omega_c^3\right){\cal R} +\nonumber\\
&&+ F^2 \Omega_c^4{\cal R} +a^2{\cal R} + \left({\eta^2\over\mu^2}+2 a F \right) \Omega_c^2{\cal R}=0~.
\ea
The dependence on $\beta_c$ has completely disappeared and one is effectively left with an ordinary differential equation for
${\cal R} \left(\Omega_c \right)$. Through the substitution, $\Omega_c = \mu z$, ${d^2\over d \Omega_c^2} = {1\over\mu^2} {d^2\over d z^2}$ and ${\cal R} \left( \Omega_c (z) \right) := \phi_a (z)$
one obtains a second order linear differential equation, which is a Schr\"odinger-like equation
\be\label{eq1.21}
- \phi_a'' (z) +V(z) \phi_a (z)= 0~,
\ee
where the potential function, V(z), reads:
\be\label{eq1.22}
V(z) = - \left(F\mu^2z^2 + \eta z-a \right)^2 + 48 \exp \left( -2 \sqrt{3} z -2 \sqrt{3} \mu^2 E z^2 + {\sqrt{3} \theta a \over\mu \lambda} \right).
\ee
Equation (\ref{eq1.21}) depends explicitly on the noncommutative parameters $\theta$, $\eta$, $\epsilon$ and the eigenvalue $a$.

\section{Thermodynamics of phase-space non-canonical noncommutative black hole}

\begin{figure}
\begin{center}
\subfigure[ ~$\eta=0$, $\theta=0$, $\epsilon=0$ and
$a=0.01$]{\includegraphics[scale=0.7]{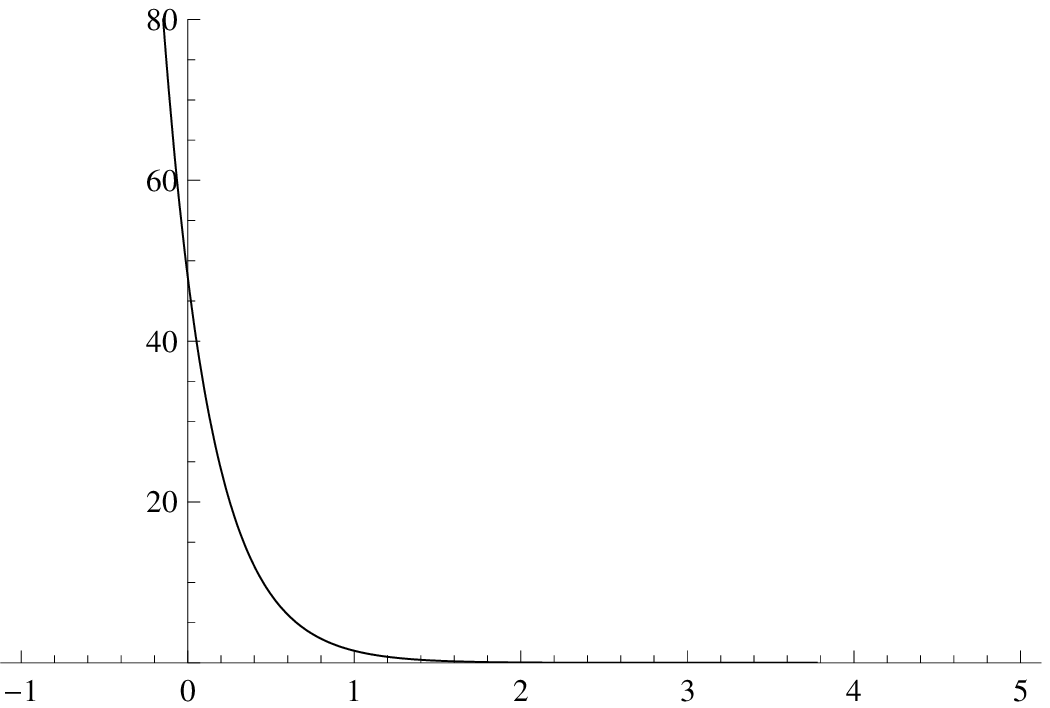}}
\subfigure[ ~$\eta=0$, $\theta=0.1$, $\epsilon=0.3$ and
$a=0.01$]{\includegraphics[scale=0.7]{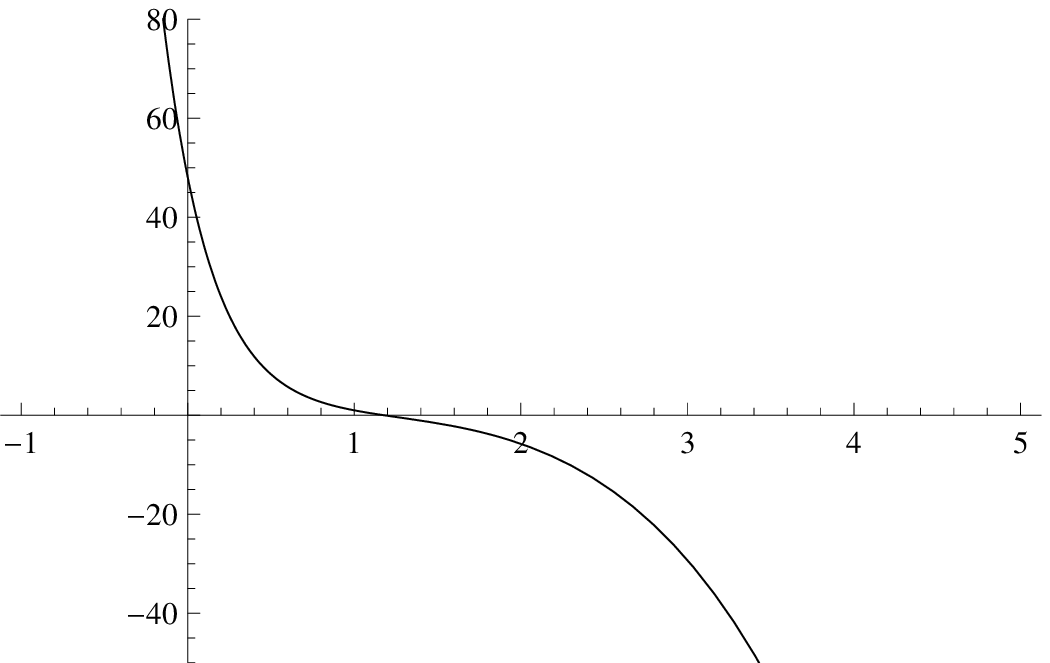}}
\subfigure[ ~$\eta=5$, $\theta=0$, $\epsilon=0.3$ and
$a=18.89$]{\includegraphics[scale=0.7]{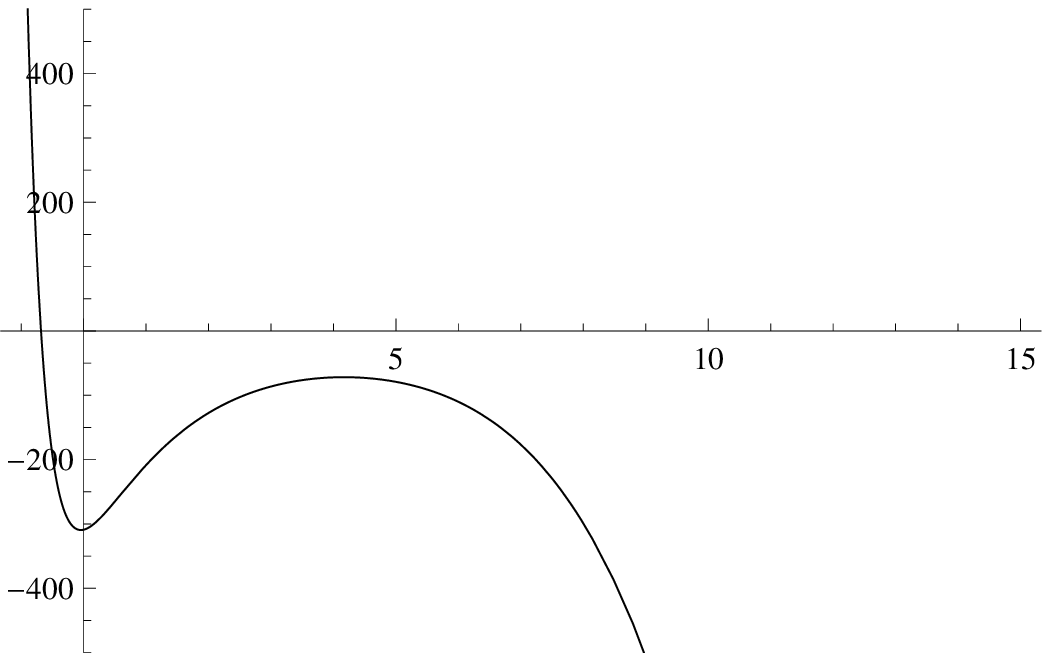}}
\subfigure[ ~$\eta=5$, $\theta=0.1$, $\epsilon=0.3$ and
$a=18.89$]{\includegraphics[scale=0.7]{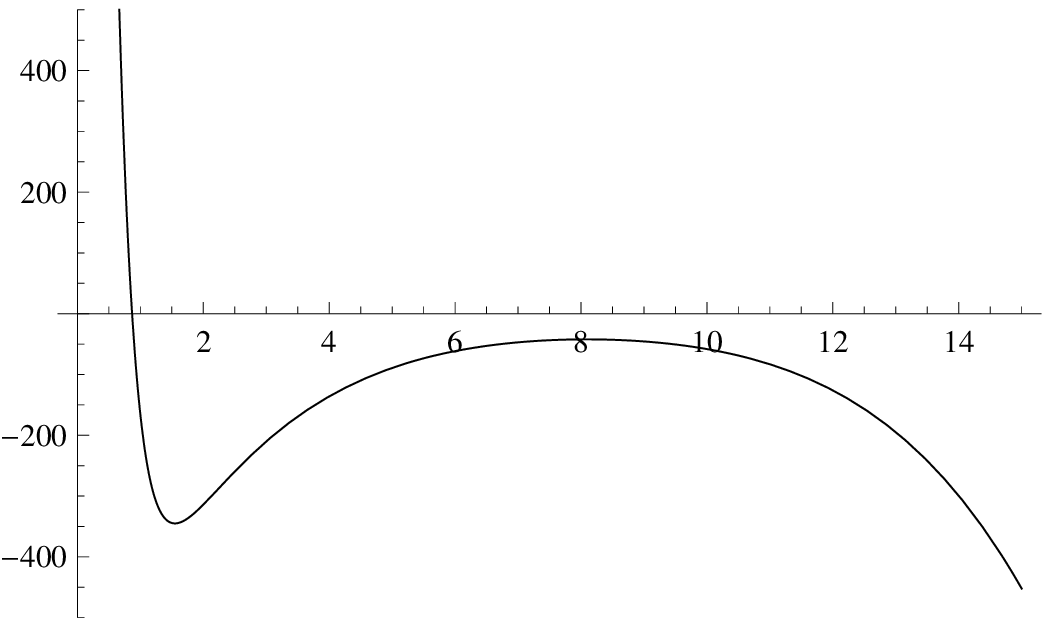}}
\caption{Potential function for some typical values of $\eta$,
$\theta$, $\epsilon$ and $a$: (a) The potential function for the
commutative case, $\theta=\eta=\epsilon=0$ and $a=0$; (b) The
potential function for the noncommutative case $\theta\neq0$ and
$\eta=0$; (c) The potential function for the momentum
noncommutative case $\eta\neq0$ and $\theta=0$; and (d) The
potential function for the full noncommutative case
$\theta,\eta\neq0$. For cases (b), (c) and (d) the non-canonical
parameter was set to $\epsilon=0.3$.} \label{fig:potential}
\end{center}
\end{figure}

In this section we compute de thermodynamical properties of the
noncommutative BH using the previous phase-space non-canonical
noncommutative quantum cosmological model. The procedure used here
is the one developed in Ref. \cite{Bastos5} for the canonical
noncommutativity. We consider the NCWDW equation to study the
quantum behaviour of the interior of the Scharwzschild BH. To
evaluate the noncommutative temperature and entropy of the BH, we
compute the partition function through the Feynman-Hibbs
procedure, applied to the minisuperspace potential function, Eq.
(\ref{eq1.22}) depicted in Fig. \ref{fig:potential}.

We present four different cases. The first case (a) describes the potential function for the commutative case $(\theta= \eta = \epsilon =0)$, a strictly decreasing exponential function without a local minimum. The second case (b), the non-canonical noncommutativity in the configuration variables $(\eta=0, \epsilon=0.3)$ is considered, which reveals a potential function also without a local extremum. Finally, in the last two cases the noncommutativity is non-canonical and imposed (c) on the momentum sector $(\theta=0)$, and (d) on the full phase-space. The last two cases admit a local minimum. Thus, as pointed out in Ref. \cite{Bastos5} the potential has a local minimum only when the noncommutativity in the momentum sector is introduced. In the three last cases ``the non-canonical" parameter was set to
$\epsilon=0.3$. This value is typical as are the values of the other noncommutative parameters. For some other values of these three parameters we encounter a qualitatively similar behavior for the
potential function. When we set $\epsilon=0$ in the potential, we recover the results of Ref. \cite{Bastos5}.

\begin{figure}
\begin{center}
\subfigure[ ~
$V(z)$]{\includegraphics[scale=0.7]{potencialNC.eps}} \subfigure[
~ $V(z)$ neglecting $F\mu^2z^2$
]{\includegraphics[scale=0.7]{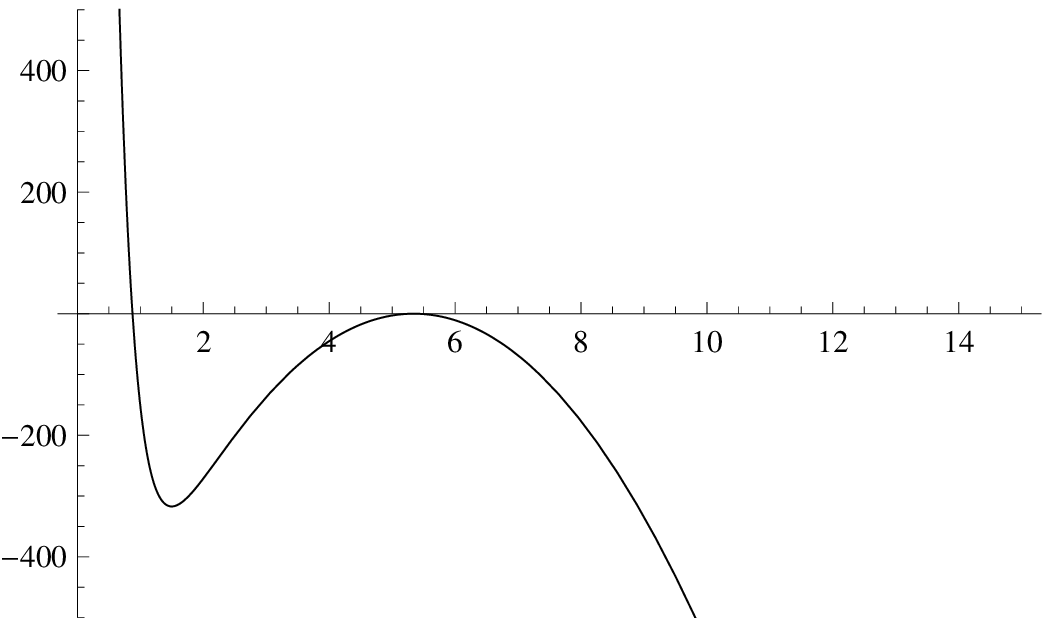}}
\caption{Approximation of the potential function. (a) the full
potential function in the non-canonical noncommutative
phase-space, $\eta=5$, $\theta=0.1$, $\epsilon=0.3$ and (b) the
potential function in the non-canonical noncommutative phase-space
neglecting the term $F\mu^2z^2$. } \label{fig:potentialapprox}
\end{center}
\end{figure}

>From Fig. \ref{fig:potentialapprox} one concludes that the potential function has a minimum, which is located at small $z$ values. Around the minimum the qualitative behavior of the two graphics (Fig. 2a) and b)) is quite similar. This allow us to safely neglect the term $F\mu^2z^2$ in the derivation of the noncommutative temperature and entropy of the BH. We stress that the Feynman-Hibbs method can be employed only when the noncommutative parameter associated with the momenta is non-vanishing. It is only under these circumstances that we have a local minimum of the minisuperspace potential. We can then expand the potential up to second order around the minimum. The potential function is, then,
\be\label{eq2.1} 
V(z)=-(\eta z-a)^2 + 48\exp{\left[-2\sqrt{3}(z- { \theta a \over 2 \mu \lambda}+\mu^2 Ez^2)\right]}~. 
\ee

In order to use the Feynman-Hibbs procedure to evaluate the partition function, one expands the exponential term in the potential Eq. (\ref{eq2.1}) to second order in powers of the $z-z_0$ variable. The minimum $z_0$ is then obtained by solving the following equation 
\be\label{eq2.2} 
{dV\over dz}|_{z_0}=\eta (\eta z_0-a)+48\sqrt{3}(1+2\mu^2Ez_0) \exp{\left[-2\sqrt{3}(z_0- {\theta a \over2 \mu \lambda}+\mu^2 E z_0^2)\right]}=0~. 
\ee 
The minimum is thus defined by the relation 
\be\label{eq2.3}
\exp{\left[-2\sqrt{3}(z_0- { \theta a \over 2 \mu \lambda}+\mu^2 E z_0^2)\right]}=-{\eta(\eta z_0-a)\over 48\sqrt{3}(1+2\mu^2Ez_0)}~.
\ee 
Moreover, it should satisfy the second derivative criterion
\be\label{eq2.4}
 6\left((1+2\mu^2Ez_0)^2-{\sqrt{3}\over3}\mu^2E\right)\exp{\left[-2\sqrt{3}(z_0- { \theta a \over 2 \mu \lambda}+\mu^2 E z_0^2)\right]}-l^2 > 0~,
\ee 
where $l:=\eta/4\sqrt{3}$. Using these relations, the second order expansion of the potential function $V(z)$ in powers of $z-z_0$ satisfies: 
\be\label{eq2.5} 
V(z)=48B(z-z_0)^2-(\eta z_0-a)^2+48k~, 
\ee 
where $B:=6k[(1+2\mu^2Ez_0)^2-(\sqrt{3}/3)\mu^2E]-l^2$ and $k:=\exp{\left[-2\sqrt{3}(z_0- {\theta a \over 2 \mu \lambda}+\mu^2 E z_0^2)\right]}$. Notice that the minimum is semiclassically stable as can be shown, following the arguments of Ref. \cite{Bastos5}.

So, we can rewrite the ordinary differential equation resulting from the NCWDW equation as
\be\label{eq2.6}
-{1\over2}{d^2\phi\over dz^2}+24\left[B-l^2\right](z-z_0)^2\phi=\left[{1\over2}(\eta z_0-a)^2-24k\right]\phi~.
\ee

Comparing Eq. (\ref{eq2.6}) with the Schr\"odinger equation of the harmonic oscillator, we can identify the noncommutative potential and the energy of the system,
\be\label{eq2.7}
V_{NC}(y)=24 (B-l^2)y^2~,
\ee
in terms of the variable $y\equiv z-z_0$. A quantum correction is introduced to the partition function through the potential, as given by the Feynman-Hibbs procedure \cite{Feynman}:
\be\label{eq2.8}
{\beta_{BH}\over24}V_{NC}''(y)=2 \beta_{BH}(B-l^2)~,
\ee
where $\beta_{BH}$ is the inverse of the BH temperature. The noncommutative potential with the corresponding quantum corrections then reads,
\be\label{eq2.9}
\mathcal{V}_{NC}(y)=24(B-l^2)\left(y^2+{\beta_{BH}\over12}\right)~.
\ee

Finally, the noncommutative partition function is given by
\be\label{eq2.10}
Z_{NC}=\sqrt{1\over{48(B-l^2)}}{1\over\beta_{BH}}\times\exp{\left[-2{\beta_{BH}^2}(B-l^2)\right]}~.
\ee
The thermodynamic quantities of the Schwarzschild BH can now be computed. Starting from the noncommutative internal energy of the BH, $U_{NC}=-{\partial\over\partial\beta_{BH}}\ln Z_{NC}$,
\be\label{eq2.11}
U_{NC}={1\over\beta_{BH}}+4(B-l^2)\beta_{BH}~,
\ee
the equality $U_{NC}=M$, allows for obtaining an expression for the BH temperature:
\be\label{eq2.12}
\beta_{BH}={M \over 8(B-l^2)}\times\left[1 \pm \sqrt{1 - {16\over M^2}(B-l^2)}\right]~.
\ee
Inverting this quantity, one obtains the BH temperature. Dropping the term proportional to $M^{-2}$, as presumably $M>>1$, and considering the positive root, we obtain:
\be\label{eq2.13}
T_{BH}={4 \over M}(B-l^2)~.
\ee
We can clearly see that this quantity is positive (cf. Eq. (\ref{eq2.4})). If $E=0$, we recover the results of Ref. \cite{Bastos5}. That is, this non-canonical noncommutativity introduces new corrections to the noncommutative temperature. Furthermore, the mass dependence of the noncommutative temperature remains the same as the Hawking temperature for a BH, $T_{BH}= {1 \over 8 \pi M}$. As in Ref. \cite{Bastos5}, we can recover the Hawking temperature for a $\eta\neq0$ value. Thus, equating Eq. (\ref{eq2.13}) with the Hawking temperature and using the stationary condition Eq. (\ref{eq2.3}), we obtain for $\theta=0.1$, $\epsilon=0.3$ and $a=18.89$:
\be\label{eq2.13a}
z_0=2.26369 \hspace{1 cm} \eta_0=0.487.
\ee
And thus, since $\eta$ cannot be exactly equal to zero, we can regard $\eta_0=0.487$ as a reference value which yields the Hawking temperature, and as $\eta$ increases we get a gradual noncommutative deformation of the Hawking temperature. Notice that this value differs from the one obtained using a canonical phase-space noncommutativity \cite{Bastos5}. This occurs as the non-canonical model has more parameters, $\epsilon$, $E$ and $F$,  it is obvious that they will affect the value of $\eta_0$.

The entropy is calculated using the expression, $S_{NC}=\ln Z_{NC}+\beta_{BH}U_{NC}$. Thus, the entropy for the phase-space noncommutative BH is given by:
\be\label{eq2.14}
S_{BH}={1\over2}+{M^2\over16B}+{M^2\over16B}\sqrt{1-{16B\over M^2}}-{1\over2}\ln{\left[{3\over2}{M^2\over B}\left(1+\sqrt{1-{16\over M^2}B}-{8B\over M^2}\right)\right]}~.
\ee
Neglecting as before terms proportional to $M^{-2}$ as $M>>1$, we finally obtain
\be\label{eq2.15}
S_{BH}\simeq{M^2\over8\left(6k[(1+2\mu^2Ez_0)^2-{\sqrt{3}\over3}\mu^2E]-l^2\right)}-{1\over2}\ln\left({3M^2\over 6k[(1+2\mu^2Ez_0)^2-{\sqrt{3}\over3}\mu^2E]-l^2}\right)~.
\ee
For the reference value $\eta=\eta_0$, we recover the Hawking entropy, plus some ``stringy" corrections:
\be
S_{BH}=4\pi M^2+\ln{\sqrt{2 \pi\over3}}-\ln({8\pi M})~.
\ee\label{eq2.16}
In summary, as in the case of the temperature, we obtain that the noncommutative entropy of the BH is the Hawking entropy plus additional contributions and some noncommutative corrections.

\section{The Singularity Problem}

Let us now turn to the singularity problem and further elaborate
on the issues discussed in Ref. \cite{Bastos3}. The wave function
Eq. (\ref{eq1.18}) solution of the NCWDW equation is oscillatory
in $\beta_c$ and the remaining $\Omega_c$-sector is
square-integrable and, hence, in the following, it is natural to
consider hyper-surfaces of constant $\beta_c$. This suggests that
a measure $d \xi = \delta ( \beta - \beta_c) d \beta d \Omega_c$
should be assumed for the definition of the inner product in $L^2
(\bkR^2)$ and the evaluation of probabilities. Starting from this
premise, we were able to prove in Ref. \cite{Bastos3} that the
probability vanishes in the vicinity of the Schwarzschild
singularity (corresponding to $\Omega_c, \beta_c \to + \infty$,
cf. Eq.(\ref{eq0.6})). Here, we wish to investigate under which
circumstances the entire KS singularity ($\beta_c \to + \infty$)
could be regularized by our non-canonical noncommutative model.

It is worthwhile reviewing our arguments. A generic solution of the NCWDW equation $\hat H \psi=0$ can be written as
\begin{equation}
\psi (\Omega_c, \beta_c) = \int C(a) \psi_a (\Omega_c, \beta_c) da~,
\label{eq2.16.1}
\end{equation}
where $C(a)$ are (for the time being) arbitrary complex constants and $\psi_a (\Omega_c, \beta_c)$ is a simultaneous solution of $\hat A \psi_a =a \psi_a$ and $\hat H \psi_a=0$,
\begin{equation}
\psi_a (\Omega_c, \beta_c)= \phi_a \left({\Omega_c\over \mu} \right) \exp \left[ \frac{i \beta_c}{\mu} \left( a - \frac{\eta}{2 \mu} \Omega_c \right)\right]~,
\label{eq2.16.2}
\end{equation}
Here $\phi_a \left(\frac{\Omega_c}{\mu} \right)$ is a solution of
Eq. (\ref{eq1.21}). This equation is asymptotically of the form
(cf. Fig. \ref{fig:potentialapprox})
\begin{equation}
-\phi_a''(z)-K z^4 \phi_a(z)=0 , \quad z \to +\infty~,
\label{eq2.16.2.1}
\end{equation}
($K$ is a positive constant) and so its solutions satisfy, cf. [pag. 75 \cite{Voronov}] ($C \in\bkR$)
\begin{equation}
\phi_a(z)\simeq \frac{C}{z} \exp \left(\pm i
\frac{K^{1/2}}{3}z^3\right) , \quad z \to +\infty ~,~ a \in \bkR~,
\label{eq2.16.2.2}
\end{equation}
and are thus square integrable. Hence we may impose the normalization:
\begin{equation}
\int \left|\psi_a (\Omega_c, \beta_c) \right|^2 d \Omega_c= \int \left| \phi_a \left(\frac{\Omega_c}{\mu} \right) \right|^2 d \Omega_c=1~.
\label{eq2.16.3}
\end{equation}
With the assumed measure, we then have:
\begin{equation}
\begin{array}{c}
|| \psi||_{L^2 (\bkR^2, d \xi)} = \left(\int \psi (\Omega_c, \beta_c) \overline{\psi (\Omega_c, \beta_c)} d \xi\right)^{\frac{1}{2}} = \\
\\
=( \int C(a) \overline{C(a^{\prime})} ( \int \psi_a (\Omega_c, \beta_c) \overline{\psi_{a^{\prime}} (\Omega_c, \beta_c)} d \Omega_c ) da da^{\prime})^{\frac{1}{2}}~.
\end{array}
\label{2.16.4}
\end{equation}
Notice that, in general, the functions $\psi_a (\Omega_c,\beta_c)$
are not orthogonal as they are solutions of a hyperbolic-type
equation (see comment in the discussion below). We may nevertheless
apply the Cauchy-Schwartz inequality to obtain:
\begin{equation}
|| \psi||_{L^2 (\bkR^2, d \xi)} \le \int | C(a)| || \psi_a  ||_{L^2 (\bkR^2, d \xi)} da = \int | C(a)| da~.
\label{eq2.16.5}
\end{equation}
We conclude that for $C(a) \in L^1 (\bkR)$ the wave function Eq.
(\ref{eq2.16.1}) belongs to $L^2 (\bkR^2, d \xi)$, i.e. it is
square-integrable on constant $\beta_c$ hypersurfaces. Hence the
probability of system reaching the Schwarzschild singularity is:
\begin{equation}
\begin{array}{c}
P(r=0,t=0) = \lim_{\widetilde{\Omega}_c, \beta_c \to + \infty} \int_{\widetilde{\Omega}_c}^{+ \infty} \int_{- \infty}^{+ \infty} \left| \psi (\Omega_c, \beta) \right|^2 d \xi = \\
\\
= \lim_{\widetilde{\Omega}_c, \beta_c \to + \infty} \int_{\widetilde{\Omega}_c}^{+ \infty}  \left| \psi (\Omega_c, \beta_c) \right|^2 d \Omega_c =0
\end{array}
\label{eq2.16.6}
\end{equation}
which effectively regularizes the singularity as point out in Ref. \cite{Bastos3}.

We next consider the more general KS singularity, corresponding to $\beta_c \to + \infty$, (c.f.  Eq.(\ref{eq0.3})). We thus wish to compute probabilities of the type:
\begin{equation}
P \left( \Omega_c \in I, \beta_c \to + \infty \right) =
 \lim_{\beta_c \to + \infty} \int_I \int_{\bkR}  | \psi (\Omega_c, \beta)|^2 d \xi = \lim_{\beta_c \to + \infty} \int_I   \left| \psi (\Omega_c, \beta_c ) \right|^2  d \Omega_c~,
\label{eq2.16.7}
\end{equation}
where $I$ is some compact subset of $\bkR$. A simple calculation shows that:
\begin{equation}
\begin{array}{c}
\left|\psi (\Omega_c, \beta_c) \right|^2 = \int_{\bkR} \int_{\bkR} C(a) \overline{C(a')} \phi_a \left(\frac{\Omega_c}{\mu} \right) \overline{\phi_{a'} \left(\frac{\Omega_c}{\mu} \right)}\exp \left(\frac{i \beta_c}{\mu} a  - \frac{i \beta_c}{\mu} a' \right) da da'=\\
\\
= \left| \int_{\bkR} C(a) \phi_a \left(\frac{\Omega_c}{\mu} \right)\exp \left(\frac{i \beta_c}{\mu} a \right) da \right|^2~.
\end{array}
\label{eq2.16.8}
\end{equation}
And thus:
\begin{equation}
P \left( \Omega_c \in I, \beta_c \to  + \infty \right) =\lim_{\beta_c \to + \infty} \int_I \left| \int_{\bkR} C(a) \phi_a \left(\frac{\Omega_c}{\mu} \right)
\exp \left(\frac{i \beta_c}{\mu} a \right) da \right|^2 d \Omega_c~.
\label{eq2.16.9}
\end{equation}
Notice that $\phi_a \in C^2 (\bkR) \cap L^2 (\bkR)$ and thus $\phi_a \in L^{\infty} (\bkR)$, that is:
\begin{equation}
M(a) \equiv ||\phi_a||_{L^{\infty} (\bkR)} = sup_{x \in \bkR} | \phi_a (x)| < \infty, \hspace{0.5 cm} \mbox{for each } a \in \bkR~.
\label{eq2.16.10}
\end{equation}
We have already assumed that $C(a) \in L^1 (\bkR)$. Given the fact
that the potential Eq. (\ref{eq1.22}) varies smoothly with respect
to $a$, it is reasonable to expect that for a suitable choice of
constants $C(a)$, we may admit the following regularity
conditions:

\vspace{0.3 cm}
\noindent
1) $C(a) \in L^1 (\bkR)$\\
2) $C(a) M(a) \in L^1 (\bkR)$

\vspace{0.3 cm}

We then have:
\begin{equation}
\int_I \left| \int_{\bkR} C(a) \phi_a \left(\frac{\Omega_c}{\mu} \right)
\exp \left(\frac{i \beta_c}{\mu} a \right) da \right|^2 d \Omega_c \le  |I|  \left[\int_{\bkR} |C(a)| M(a)da \right]^2 < \infty~,
\label{eq2.16.11}
\end{equation}
where we have used the regularity condition 2), and where $|I|= \int_I d \Omega_c < \infty$, since $I$ is compact. 
We thus have a uniform convergence on every compact interval $I$ and all $\beta_c$. This means that we can safely interchange the integral and the limit to obtain:
\begin{equation}
P \left( \Omega_c \in I, \beta_c \to + \infty \right) = \int_I \lim_{\beta_c \to + \infty} \left| \int_{\bkR} C(a) \phi_a \left(\frac{\Omega_c}{\mu} \right)
\exp \left(\frac{i \beta_c}{\mu} a \right) da \right|^2 d \Omega_c~.
\label{eq2.16.12}
\end{equation}
>From the regularity condition 2) the functions $C(a) \phi_a
\left(\frac{\Omega_c}{\mu} \right)$ regarded as functions of $a$
(for each fixed $\Omega_c$) belong to $L^1 (\bkR)$. Indeed, $|C(a)
\phi_a \left(\frac{\Omega_c}{\mu} \right)| \le |C(a) M(a)| $, for
all $\Omega_c \in \bkR$. On the other hand, the Riemann-Lebesgue
Lemma \cite{Stein} states that every $L^1 $ function has a
continuous Fourier transform, which vanishes at infinity, i.e.:
\begin{equation}
\lim_{|\beta_c| \to \infty} \int_{\bkR} f(a) e^{\frac{i \beta_c}{\mu} a } da =0
\label{eq2.16.13}
\end{equation}
for any $f \in L^1 (\bkR)$.
We conclude that:
\begin{equation}
P \left( \Omega_c \in I, \beta_c \to + \infty \right) =0
\label{eq2.16.14}
\end{equation}
for any compact set $I$. This effectively regularizes the KS singularity as well.

\section{Discussion and Conclusions}

In this work we obtain the noncommutative temperature and entropy
of the Schwarzschild BH using a non-canonical noncommutative
extension of a KS cosmological model. This formulation properly
generalizes the full commutative, as well as the configurational
and the momentum (canonical) noncommutative, models. We showed
that the use of the Feynman-Hibbs method is only meaningful if one
admits a non-vanishing noncommutativity in the momentum sector
(i.e. $\eta \not=0$) and we find that the non-canonical
noncommutativity introduces further corrections to the
noncommutative temperature and entropy obtained in Ref.
\cite{Bastos5}. Just as in Ref. \cite{Bastos5}, the Hawking
quantities for the BH are recovered for a non-vanishing value of
$\eta_0$, which in the present setup corresponds to
$\eta_0=0.487$.

In the second part of the paper we have showed that the KS
singularity (just as the Schwarzschild singularity, studied in a
previous paper) is regularized by the non-canonical,
noncommutative regime. This is the main feature of our model and a
direct consequence of the fact that the eigenstates $\phi_a$ are
square integrable, in spite of being associated to a continuous
set of eigenvalues $a$. This property is shared by all
Hamiltonians of the form $H=p^2+V(x)$ with a potential
asymptotically like $V(z) \simeq -Kz^{2+\epsilon}$ (for some
$K,\epsilon>0$) as $z \to +\infty$ \cite{Voronov}. The reason for
the apparent paradox of having square integrable eigenstates and a
continuous spectrum is that the Hamiltonians $H$ defined on their
maximal domain
\begin{equation} \label{eq.3.1}
D_{max}(H)=\{\psi \in L^2(\bkR): H\psi \in L^2(\bkR) \}
\end{equation}
are not self-adjoint operators. In fact, they are the adjoint operators of
the symmetric Hamiltonians
\begin{equation}
H^{(0)}:{\cal S}(\bkR) \to L^2(\bkR) , \, \phi \to H^{(0)}
\phi=(p^2+V)\phi \label{eq3.2}
\end{equation}
(where ${\cal S}(\bkR)$ is the set of Schwartz functions on
$\bkR$) and act on domains larger than the domains of the
self-adjoint extensions of $H^{(0)}$. Since $H$ is not
self-adjoint, it is not surprising it generates a non-unitary time
evolution.

It is worthwhile to compare these results with the full
commutative case ($\theta=\eta=\epsilon=0$) and the momentum
noncommutative case ($\theta=\epsilon=0$, $\eta\not=0$). In the
full commutative case (as also for configurational
noncommutativity, i.e. $\eta=\epsilon=0$ but $\theta \not=0$) the
potential Eq.(\ref{eq1.22}) behaves asymptotically as $V(z) \simeq
-a^2$. Consequently, the Hamiltonian Eq.(\ref{eq3.2}) has
deficiency indices $(0,0)$, i.e. it is essentially self-adjoint, and
its (unique) self-adjoint realization is defined on the maximal
domain Eq.(\ref{eq.3.1}). Hence, its spectrum is continuous, its
eigenstates are not normalizable and the time evolution is
unitary.

Most of these properties are also shared by the momentum
(canonical) noncommutative case, where the potential behaves
asymptotically like $V(z) \simeq -z^2$. The operator $H$, defined
on the domain Eq.(\ref{eq.3.1}), is still self-adjoint. Its
eigenstates display a pronounced damping behavior which,
nevertheless, is not sufficient to make them normalizable. The
spectrum is continuous and the evolution unitary \cite{Bastos5}.

Once we give up the self-adjointness of the Hamiltonian, we may
become apprehensive that the corresponding eigenvalues may not be
real. We stress that a self-adjoint operator is a sufficient
condition for the reality of the eigenvalues. However, it is by no
means necessary. In fact, a wide class of non self-adjoint
Hamiltonian operators has recently been studied which exhibit real
eigenvalues \cite{Bender}. These Hamiltonians are $PT$-symmetric,
that is they display an invariance under parity and time reversal
transformations. An unbroken $PT$-symmetry can be shown to be
equally sufficient for a real spectrum. Our Hamiltonian in Eqs.
(\ref{eq1.21}) and (\ref{eq1.22}) looks asymptotically like
$p^2-Kx^4$ (with $K>0$). This Hamiltonian has been analyzed in
\cite{Bender} and shown to be $PT$-symmetric.

\subsection*{Acknowledgments}

\vspace{0.3cm}

\noindent The work of CB is supported by Funda\c{c}\~{a}o para a
Ci\^{e}ncia e a Tecnologia (FCT) under the grant
SFRH/BPD/62861/2009. The work of OB is partially supported by the
FCT grant PTDC/FIS/111362/2009. NCD and JNP were partially
supported by the grants PTDC/MAT/69635/2006 and
PTDC/MAT/099880/2008 of FCT.

\end{document}